\documentclass[reprint,amsmath,amssymb,aps,prl,longbibliography]{revtex4-2}
\usepackage{graphicx}
\usepackage{amsmath}
\usepackage{amssymb}
\usepackage{color}
\begin{document}
\title{Magnetoelectric imprint of skyrmions in van der Waals bilayers}
\author{Zhong Shen}
\author{Xiaoyan Yao}
\email[]{yaoxiaoyan@seu.edu.cn}
\author{Shuai Dong}
\email[]{sdong@seu.edu.cn}
\affiliation{Key Laboratory of Quantum Materials and Devices of Ministry of Education, School of Physics, Southeast University, Nanjing 211189, China}
\date{\today}

\begin{abstract}
To effectively track and manipulate topological solitons (e.g. skyrmions) are the key challenge before their applications. Inspired by the idea of sliding ferroelectricity, here a general strategy is proposed to print magnetic skyrmions to electric skyrmions in van der Waals bilayers. Through the proximate interactions, there is an isoperiodic bijection relationship between local dipoles and spin moments. This magnetoelectric imprint effect not only extends the strategies to create electric skyrmions, but also leads to an approach for all-electrical readout/manipulation of magnetic skyrmions.
\end{abstract}
\maketitle

\textit{Introduction}.
Magnetic (M) skyrmions are vortex-like spin textures with integral topological numbers, which open a promising direction for spintronics~\cite{Nagaosa2013NN,Fert2017NRM}. Since the experimental observations of M-skyrmions in reciprocal space~\cite{Muhlbauer2009S} and real space~\cite{Tokura2010N}, the physical mechanisms of their creation and manipulation have been extensively studied, and more and more materials with M-skyrmions have been predicted/found~\cite{Shen2023PRB,Yao2022PRB,Yao2023i,Shen2022PRB,Kanazawa2021CR,Amoroso2020NC,Zhang2019PRB}. Up to now, M-skyrmions can be directly visualized by Lorentz transmission electron microscopy~\cite{Tokura2010N} and spin-polarized scanning tunnelling microscopy~\cite{Romming2013S,Wiesendanger2015PRL}, which sets a high technical threshold for experimental direct characterizations. Thus, other easier experimental routes are urgently needed to track and manipulate M-skyrmions directly. Although the electrical readout of M-skyrmions was attempted, such as spin-mixing magnetoresistance (XMR)~\cite{Lounis2015NC,Hanneken2015NN} and chiral XMR (C-XMR)~\cite{LimaFernandes2022NC} based on current signals, and the dielectric/ferroelectric responses of M-skyrmion based on electric field signals. The Joule heat is inevitable in magnetoresistance measurements while the dielectric/ferroelectric responses only works in insulators~\cite{Chu2015SR,Okamura2013NC,Mochizuki2013PRB,Yao2020NJP,Xia2023PRL}.

The electric (E) skyrmions, with similar topological textures of electric dipoles, were also proposed and experimentally realized in PbTiO$_3$/SrTiO$_3$ superlattices~\cite{Das2019N}. Later, the concept of E-skyrmions/vortex has been generalized to more systems ~\cite{Zhu2022PRL,Yuan2023PRL,Shao2023NC,McCarter2022PRL,Wang2024NC,Yang2021NC}. Novel physics, e.g., large second harmonic generation (SHG), negative capacitance, and emergent chirality, have been reported for E-skyrmions~\cite{Ramesh2021NM,Salahuddin2019N,Shao2023NC}. Different from the M-skyrmions, the E-skyrmions can be directly regulated by electric field, which is much easier and more energy-efficient. However, current E-skyrmions are mostly limited in titanates and their superlattices~\cite{Zhu2022PRL,Yuan2023PRL,Shao2023NC,McCarter2022PRL,Wang2024NC}. Usually, the complex interplay among elastic, electrostatic, and gradient energy is required to stabilize the E-skyrmions~\cite{Shimada2024PRL}, namely the origin of E-skyrmions is mostly related to the electric-mechanical coupling. It is as expected since in these systems the electric dipoles are from structural distortions.

In some new kinds of ferroelectrics, the distortion of electron clouds can also contribute a lot to the local dipoles and macroscopic polarizations. For example, in the type-II multiferroic o-HoMnO$_3$ and Hf$_2$VC$_2$F$_2$, the magnetism-induced polarizations can be mostly from the distortion of electron clouds, instead of the structural distortions~\cite{Picozzi2007PRL,Zhang2018JotACS}. Another case is the sliding ferroelectrics in van der Waals (vdW) materials~\cite{Li2017AN,Wu2021PNAS}, in which the polarization originates from the bias of electron clouds due to the interlayer interaction~\cite{Ding2021PRM}. Comparing with those conventional dipoles from structural distortions, these electronic-originated dipoles are naturally advantageous for ultra-high-speed switching since the dynamics of electrons are much faster than ions. Thus, it will be highly interesting and vital to pursue E-skyrmions with dipoles from electron clouds distortion.

In this Letter, we propose a mechanism to generate E-skyrmions by rubbing M-skyrmions to adjacent ferromagnetic layer, which is coined as magnetoelectric imprint (MEI). Microscopically, the dipoles come from the electron clouds distortion, which is spin-orientation dependent.  Based on this MEI effect, M-skyrmions can be read/manipulated using pure electrical field methods. Our work is mainly based on first-principles density of functional theory (DFT) calculations and atomistic simulation of spin lattice model, and more details can be found in Supplemental Materials (SM)~\cite{SM}.

\begin{figure}
\centering
\includegraphics[width=0.4\textwidth]{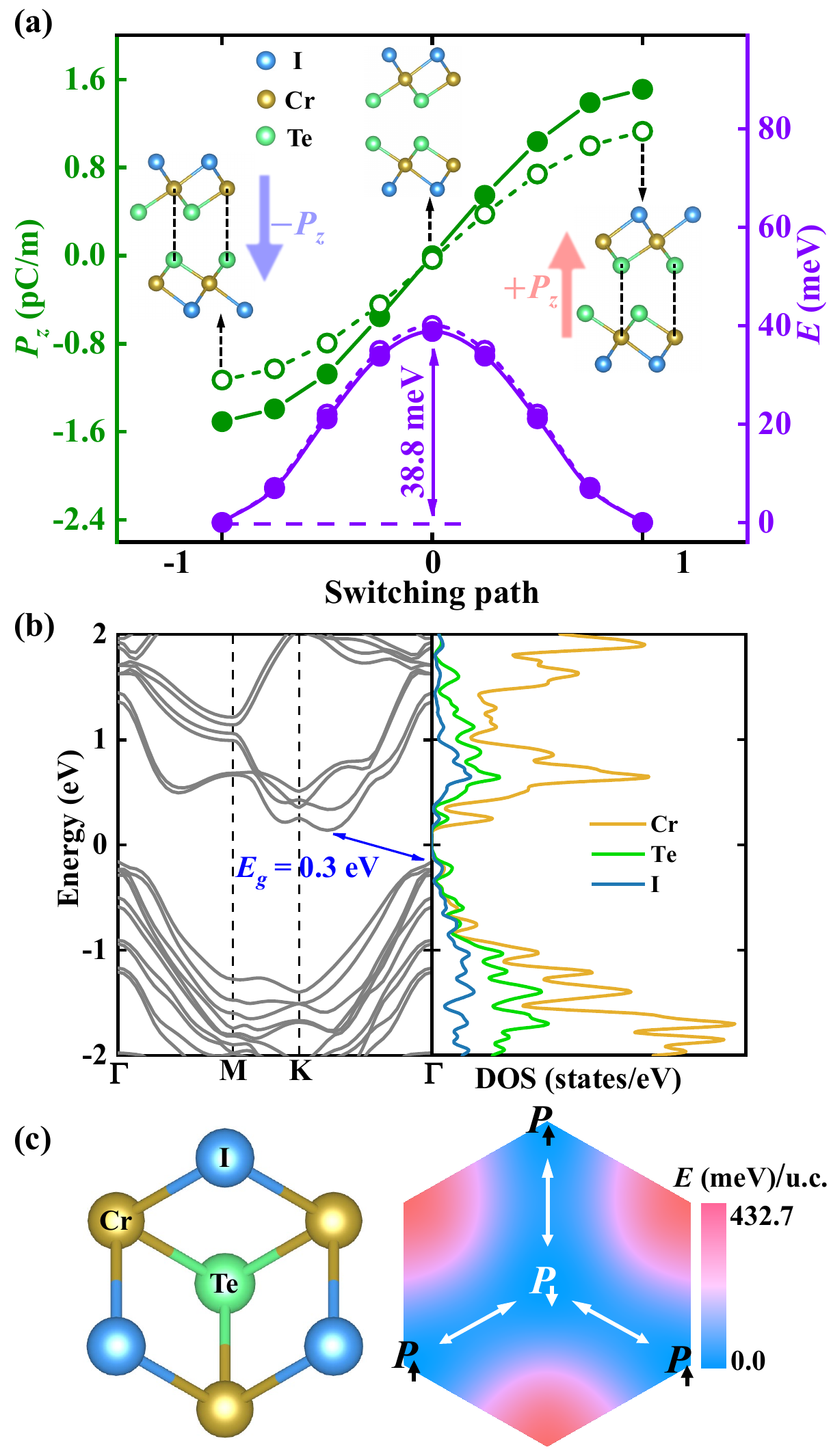}
\caption{Physical properties of CrTeI bilayer. (a) The off-plane ferroelectric polarization (green dots) and switching energy barrier (purple dots) along possible switching path, with A-type antiferromagnetic order. Open symbols: the polarization and barrier obtained with ferromagnetic order. Insets: the side views of ferroelectric and paraelectric states. (b) Electronic structure. Left: band structure; Right: element-projected density of states (DOS). (c) Left: the top view of CrTeI. Right: the energy map for sliding.}
\label{fig1}
\end{figure}

\textit{Sliding magnetic bilayers}. 
Our model system is CrTeI bilayer, as shown in Fig.~\ref{fig1}(a). In each monolayer, the magnetic Cr cation is sandwiched by two sheets of nonmagnetic anions (I and Te), forming a hexagonal lattice with a point group $C_{3v}$. We chose CrTeI for the following reasons. First, the Janus monolayer naturally breaks the spatial inversion symmetry along the off-plane direction, which provides an inborn Dzyaloshinskii-Moriya interaction (DMI), a key driving force for M-skyrmions~\cite{Dzyaloshinsky1958JPCS,Moriya1960PR}. Second, the heavy elements (Te and I) are advantageous to obtain strong spin-orbit coupling (SOC), guaranting a large DMI in this Janus monolayer.  Last, for practical consideration, similar Janus structure, e.g. MoSSe~\cite{Lu2017NN,Zhang2017AN}, has been synthesized experimentally.

Considering the asymmetric upper/lower surfaces of Janus structure, there are at least three kinds of stacking sequences for CrTeI bilayer: 1) Te-I---Te-I, 2) Te-I---I-Te, and 3) I-Te---Te-I. As compared in Fig.~S1 in SM~\cite{SM}, the I-Te---Te-I stacking sequence owns the lowest energy, which will be studied in the following. Then the most stable structures, i.e. the $\pm P_z$ states in Fig.~\ref{fig1}(a), are obtained by relaxing the structure from different initial structures (Fig.~S2 in SM~\cite{SM}). And the dynamic stability of the $\pm P_z$ structures is confirmed by the phonon dispersion spectrum [Fig.~S2(m) in SM~\cite{SM}]. 

As demonstrated in Fig.~S3 in SM~\cite{SM}, the interlayer magnetic coupling is found to be antiferromagnetic, while the intralayer coupling is ferromagnetic. Thus the ground state is the A-type antiferromagnetism. The magnetocrystalline anisotropy is also calculated, which supports a magnetic easy axis along the off-plane direction.

Then the electronic structure is calculated based on the GGA+$U$ method ($U_{\rm eff}=1.5$ eV for Cr's $3d$ orbitals) with SOC, for the A-type antiferromagnetic state. An indirect band gap is obtained, as shown in Fig.~\ref{fig1}(b). It is well known that DFT systematically underestimates band gaps. For supplementary, the Heyd-Scuseria-Ernzerhof (HSE06) hybrid functional~\cite{Heyd2003TJoCP,Heyd2006TJoCP} was also tested, which leads to a larger band gap $0.8$ eV [Fig.~S3(d) in SM]~\cite{SM}. 

Although the I-Te---Te-I stacking sequence seems to be centrosymmetric, off-plane polarizations ($\pm P_z$) can emerge from the interlayer sliding, i.e. the sliding ferroelectricity. As shown in Fig.~\ref{fig1}(a), the bistable $+P_z$ and $-P_z$ states can be switched via sliding. The amplitude of $P_z$ reaches $1.51$ pC/m, which is almost an order of magnitude larger than that of $T_d$-WTe$_2$ ($0.2$ pC/m)~\cite{Fei2018N,Yang2018JPCL}, and comparable to those of heterobilayer MoS$_2$/WS$_2$ ($1.45$ pC/m)~\cite{Rogee2022S} and bilayer $h$-BN ($1.88$ pC/m)~\cite{Yasuda2021S,Hod2021S}. This sliding-dependent $P_z$ is mostly SOC-independent.  And the switching barrier is moderate $38.8$ meV/u.c., close to the case of ZrI$_2$~\cite{Ding2021PRM}.  According to the energy landscape [Fig.~\ref{fig1}(c)], this interlayer sliding direction is along the projection direction of I-Te bonds. 

Above results are obtained with the A-type antiferromagnetic condition. By using the ferromagnetic conditions, above conclusion of sliding ferroelectricity and its switching barrier remain robust, although the amplutide of polarization will be changed accordingly [open symbols in Fig.~\ref{fig1}(a)]. In other words, the interlayer magnetic order can quantitatively tune the sliding ferroelectric polarization in CrTeI bilayer, implying a direct magnetoelectricity. Similar behavior was also found in CrI$_3$/MnSe$_2$ vdW heterostructure~\cite{Kan2023arXiv}.

\begin{figure}
\centering
\includegraphics[width=0.48\textwidth]{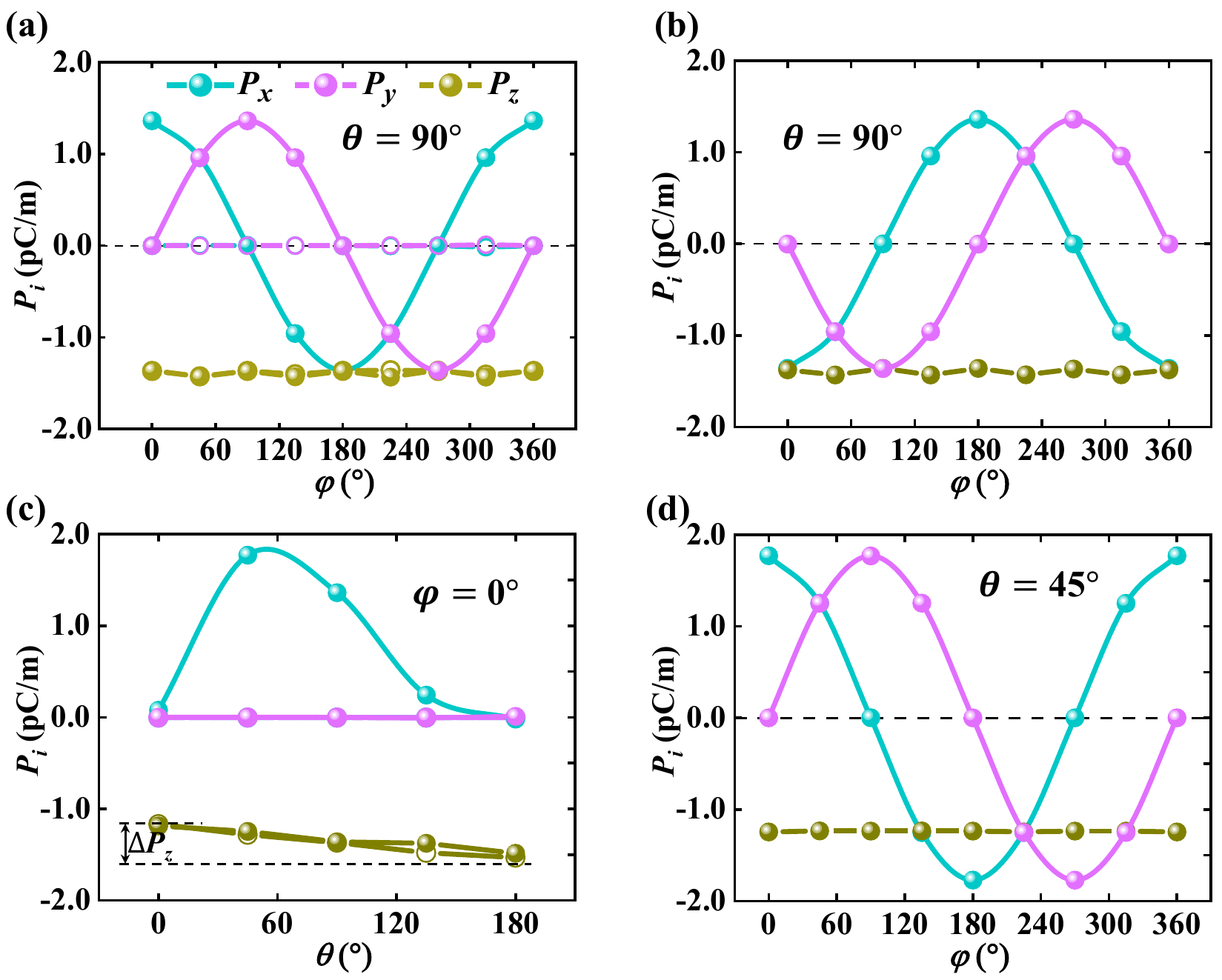}
\caption{Spin-dependent polarization ($P_x$, $P_y$, $P_z$) in CrTeI bilayer. Here $\theta$ and $\varphi$ are the polar and azimuthal angles of the free-layer spins $\textbf{S}$, while the frozen-layer spins are fixed along $+z$ [or $-z$ in (b)]. (a) $\theta=90^\circ$. The evolution of ($P_x$, $P_y$) follows ($S_x$, $S_y$) synchronously.  For comparison, ($P_x$, $P_y$) disappears when SOC is switched off (open symbols). (b) The same as (a) but the frozen-layer spins are fixed along $-z$. (c) $\varphi=0^\circ$. (d) $\theta = 45^\circ$.}
\label{fig2}
\end{figure}

\textit{Magnetoelectric imprint}.
Above direct magnetoelectricity is based on collinear magnetic orders, which can be extended to more general cases, i.e. interlayer noncollinear configurations. To study this effect, all spins in the bottom layer are fixed ferromagnetically along the off-plane $+z$ direction if not noted explicitly, denoted as the frozen layer. Then the orientation of spins in the top layer is rotated as a whole (denoted as the free layer), characterized by its polar angle $\theta$ and azimuthal angle $\varphi$.

First, by keeping $\theta=90^\circ$ in the free layer, the evolution of polarization is calculated as a function of $\varphi$. As shown in Fig.~\ref{fig2}(a), the $z$-component of polarization ($P_z$) is almost a constant, while the in-plane components ($P_x$ and $P_y$) change significantly and periodically, namely the vector $\textbf{P}_{xy}$ synchronously rotates following the spins in the free layer, i.e. $\textbf{P}_{xy}||\textbf{S}_{xy}$.  Coincidentally, the amplitudes of spin-induced $P_{xy}$ and the sliding-induced $P_z$ are almost identical: $\sim1.4$ pC/m. Thus, the vector of total ferroelectric polarization $\textbf{P}$ forms a conical trajectory with solid angle $\sim45^\circ$.

Second, this spin-dependent $\textbf{P}_{xy}$ originates from SOC, which is zero without SOC [open symbols in Fig.~\ref{fig2}(a)]. If the spins in the frozen layer are fixed to $-z$, the induced $\textbf{P}_{xy}$ becomes antiparallel to the spins' orientation in the free layer, i.e. $-\textbf{P}_{xy}||\textbf{S}_{xy}$, as shown in Fig.~\ref{fig2}(b).

Third, the polar angle $\theta$ is tuned with fixed $\varphi=0^\circ$, which leads to a moderate change of $P_z$ ($\Delta P_z\approx0.4$ pC/m $\sim1/3P_z$). Meanwhile, $P_x$ changes significantly within one period and $P_y$ is zero, as shown in Fig.~\ref{fig2}(c). Noting that $P_x$($\theta$) is asymmetry with respect to $\theta=90^\circ$, due to the broken mirror symmetry caused by interlayer sliding and the spins of frozen layer. Further, $\Delta P_z$ becomes opposite once the fixed spins in bottom layer is reversed to $-z$ (not shown here). Different from $\textbf{P}_{xy}$, $\Delta P_z$ is not SOC-originated [see open symbols in Fig.~\ref{fig2}(c) for the non-SOC results]. For completeness, more evolutions of $\theta$-dependent $\textbf{P}$ for other fixed $\varphi$'s are shown in Fig.~S4 in SM~\cite{SM}.

Fourth, the $\varphi$-dependent $P_x$/$P_y$/$P_z$ component for $\theta=45^\circ$ are shown in Fig.~\ref{fig2}(d), which is qualitatively  identical to the $\theta=90^\circ$ case.

According to aforementioned results, the vector $\textbf{P}$ can be phenomenologically expressed as: 
\begin{equation}
\textbf{P} = P_0\hat{z} + (\textbf{S}_f \cdot \hat{z})(aS_x\hat{x} + aS_y\hat{y} + bS_z\hat{z}),
\end{equation}
where $\textbf{S}$ ($\textbf{S}_f$) is the normalized spin vector of the free (frozen) layer and $P_0$ is the base line of sliding polarization. The signs of coefficients $a$/$b$ are determined by the frozen layer spins, as an indication of interlayer coupling and requested by the time-reversal symmetry of $\textbf{P}$. Most importantly, there is a bijection between $\textbf{P}$ and $\textbf{S}$, i.e. one-to-one correspondence, coined as the MEI here. In the microscopic level, such bijection relies on the hybrid mechanisms of exchange strictions~\cite{Sergienko2006PRL} and KNB model ~\cite{Katsura2005PRL} (Fig.~S5 in SM \cite{SM}).

\begin{figure*}
\centering
\includegraphics[width=0.8\textwidth]{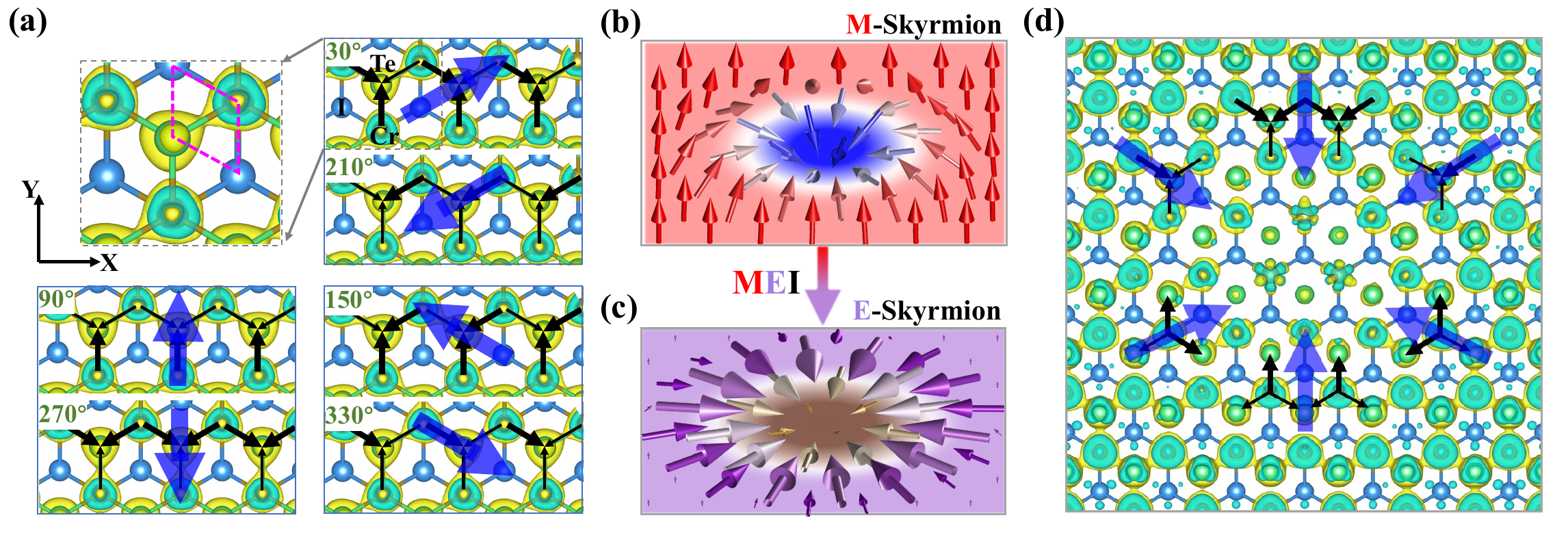}
\caption{Visualization of magnetoelectric imprint. (a) Image of differential electron density (DED) of the frozen layer as a function of uniform spin orientation of free layer. Here $\theta=90^\circ$ without loss of generality.  The DED is defined as $\delta\rho(\textbf{r})=\rho_S(\textbf{r})-\rho_0(\textbf{r})$, where $\rho_S(\textbf{r})$ is the electron density at position $\textbf{r}$ for giving $\textbf{S}$ of the top layer and $\rho_0(\textbf{r})$ is the ground state one. Black arrows: local dipoles of Cr-Te bonds  defined as $\textbf{d}=-e\int_V\delta\rho(\textbf{r})d\textbf{r}$, where $e$ is the elementary charge and $V$ is the volume of each bond (pink rhombus).  Blue arrows: the net polarization. (b) The M-skyrmion in the free layer used in DFT calculation. (c) The E-skyrmion generated in the frozen layer, derived via the bijection relation. (d) The corresponding DED of the frozen layer obtained from DFT, which shows the evidence of an E-skyrmion.}
\label{fig3}
\end{figure*}

Then it is interesting to ask whether the MEI function can persist when there is no sliding between CrTeI bilayers. As shown in Fig.~S6 in SM \cite{SM}, for the nonpolar stacking bilayer (space group $P\overline{6}m2$), the in-plane component $\textbf{P}_{xy}$ can still follow $\textbf{S}$ of the top layer, with reduced amplitude ($\sim50\%$ of $\textbf{P}_{xy}$). However, the $P_z$ component becomes negligible, despite the orientation of $\textbf{S}$. In other words, the breaking reversal symmetry along the $z$-axis is a precondition to the full MEI function.

The magnetism induced $\textbf{P}$ can be visualized using the differential electron density (DED) of the frozen layer. As shown in Fig.~\ref{fig3}(a), the three Cr-Te bonds are no longer equivalent when $\textbf{S}_{xy}$ is nonzero, breaking the in-plane $C_3$ rotational symmetry (See Fig.~S7 for more details). Then a local electric dipole can be generated for each bond. These bond dipoles are shown as black arrows in Fig.~\ref{fig3}(a) and the macroscopic $\textbf{P}_{xy}$ (blue arrow) is the superposition of these dipoles. 

This MEI effect can be further extended to non-uniform spin textures, e.g. M-skyrmions. According to the aforementioned MEI rule, i.e. the isoperiodic bijection between $\textbf{P}$ and $\textbf{S}$, the generated dipoles should also be non-uniform, which form an E-skyrmions, as shown in Fig.~\ref{fig3}(b-c). To verify this expectation, the DFT calculation is performed using a supercell (containing $112$ Cr ions) with a M-skyrmion in the free layer. Then the E-skyrmion of ($P_x$, $P_y$, $\Delta P_z$) can be clearly evidenced in the ferromagnetic frozen layer via the DED image, as shown in Fig.~\ref{fig3}(d). In this sense, the M-skyrmions in one layer can be detected (and even manipulated) in its proximate layer via pure electrical methods.

\textit{Conditions for magnetoelectric skyrmions}. 
Then the next question is how to stablize a M-skyrmion in one layer while the rest layer keeps ferromagnetic, since the bilayer are ``identical" in the chemical component. Luckily, the off-plane polarization from sliding breaks the symmetry between these two layers, openning a route to pursuit hetero-magnetism in such a bilayer. So far, the twofold role of sliding ferroelectricity is clear: 1) break the reversal symmetry along $z$-axis, which is a precondition to the full MEI function. 2) create ideal conditions for pursuing hetero-magnetism in homostructures (e.g. CrTeI bilayer).

\begin{figure}
\centering 
\includegraphics[width=0.47\textwidth]{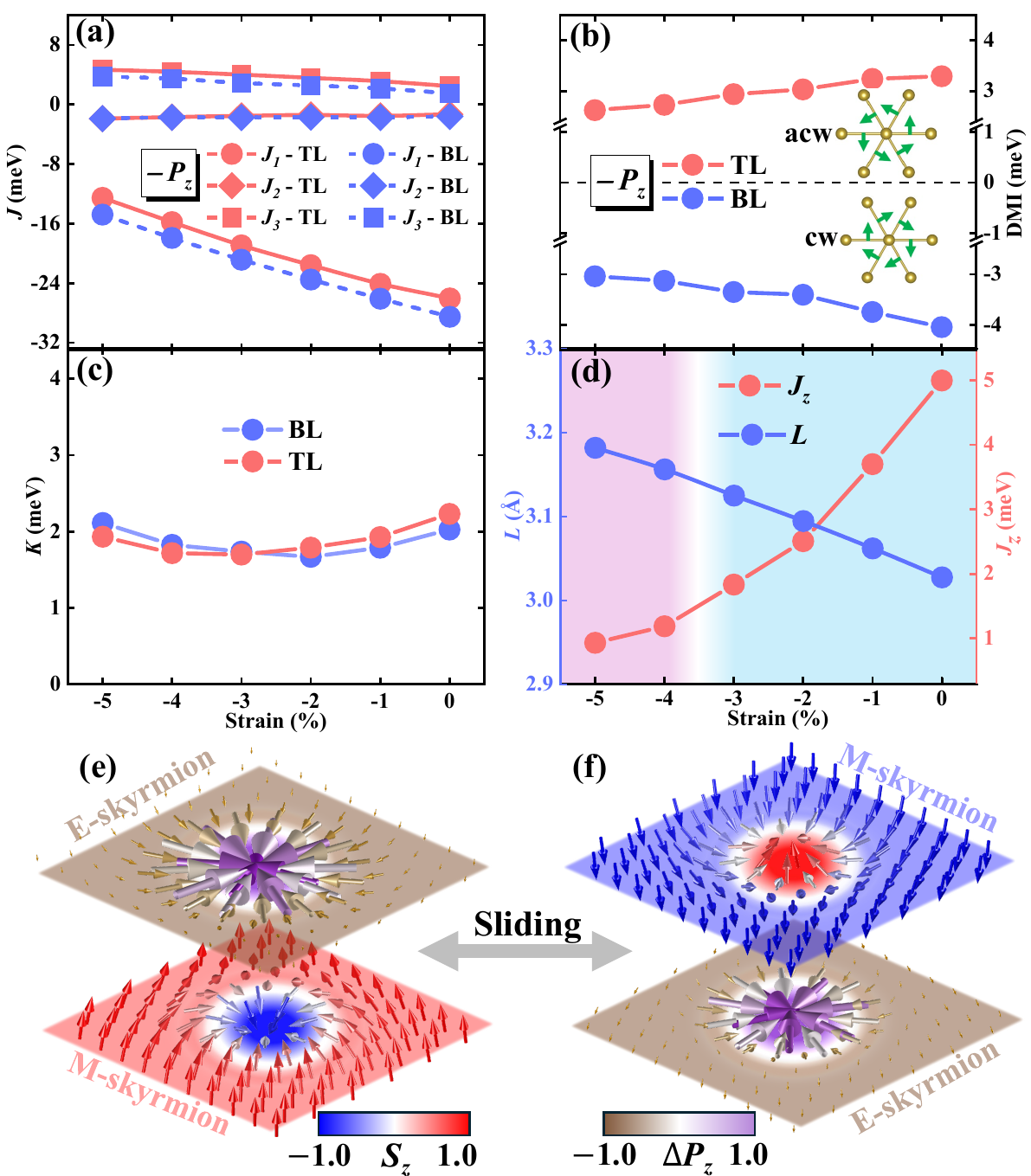}
\caption{Layer-resolved magnetic coefficients under biaxial compressive strain and M/E-skyrmions. (a) The Heisenberg-type exchange coefficients $J_i$ ($i$: the $1$st/$2$nd/$3$rd-nearest-neighbor index). (b) The nearest-neighboring DMI. The signs of DMI of top/bottom layers are opposite, which prefer opposite chirality (clockwise \textit{vs} counterclockwise).  (c) The single-ion magnetocrystalline anisotropy. (d) The interlayer distance $L$ and exchange coefficient ($J_z$). `TL'/`BL' represents the top/bottom layer, respectively. According to the model simulation, isolated skyrmion can appear in the strained regions [magenta in (d)]. (e-f) The model simulated spin/dipole textures under $-4\%$ strain for $-P_z$ and $+P_z$ states, respectively.}
\label{fig4}
\end{figure}

To demonstrate this effect, the layer-resolved magnetic interactions are calculated as functions of in-plane biaxial strain, as shown in Fig.~\ref{fig4} [see Figs.~S8-S9 in SM for more details \cite{SM}]. With increasing compressive strain, the nearest-neighboring ferromagnetic exchange $J_1$ is significantly suppressed (e.g. $\sim50\%$ upon $-5\%$ strain) but the DMI ($D$) is only moderately suppressed (e.g. $\sim25\%$ upon $-5\%$ strain). Meanwhile, the magnetocrystalline anisotropy $K$ is insensitive to the strain. In this sense, the $|D/J_1|$ ratio becomes larger under compressive strain, which is advantageous to stabilize M-skyrmions. 

Furthermore, the exchanges are indeed different between two layers, as shown in Fig.~\ref{fig4}(a). For the $-P_z$ case, comparing with the bottom layer, $J_1$ is weaker for $\sim8-15\%$ in the top layer. Meanwhile, the amplitude of $D$ in the top layer is even smaller for $\sim11-19\%$ [Fig.~\ref{fig4}(b)]. Thus the larger $|D/J_1|$ ratio in the bottom layer is in favour of M-skyrmions. In addition, the compressive strain enlarges the vdW gap ($L$) from $3.03$ \AA{} to $3.18$ \AA, which seriously reduces the interlayer exchange $J_z$ by $80\%$ [Fig.~\ref{fig4}(d)], which is also advantageous to decouple the magnetic textures between two layers.

By using these coefficients, an atomistic simulation is performed on a spin lattice model \cite{SM}. An isolated M-skyrmion with diameter $\sim4$ nm can be indeed obtained in the bottom layer when the compressive strain reaches $-4\%$ [Fig.~\ref{fig4}(e)]. Meanwhile, an E-skyrmion is induced in the ferromagnetic top layer, as a consequence of aforementioned MEI effect. More interestingly, by sliding one layer to switch to the $+P_z$ state, the roles of top and bottom layers are reversed [see Fig.~S10 for $J$ and DMI]. Then the M-skyrmion can appear in the top layer while the corresponding E-skyrmion is generated in the bottom layer [Fig.~\ref{fig4}(f)].The MEI of skyrmion at $-4\%$ strain is also confirmed by directly DFT calculations (Fig.~S11 in SM)~\cite{SM}.

More importantly, such MEI effect is robust and general in other vdW (homo)hetero-structures as demonstrated in Fig.~S12-S13, and potential challenges related to the stability and practicality are discussed in SM~\cite{SM}.

In summary, a general strategy called magnetoelectric imprint has been proposed, which can map M-skyrmions to E-skyrmions in the proximate layer via interlayer magnetic couplings and spin-orbit coupling. This kind of E-skyrmions is generated via electronic cloud distortion instead of lattice distortions. Furthermore, such a pair of M-skyrmion and E-skyrmion can be viewed as an emerging quasiparticle (i.e. ME-skyrmion), which provides the opportunity for full electrical field characterization of magnetic topological textures.

\begin{acknowledgments}
We are grateful to Haoshen Ye, Ziwen Wang, Dr. Ning Ding, Yu Xing, Xiong Luo, and Dr. Linglong Li for helpful discussions. This work is supported by National Natural Science Foundation of China (Grant Nos. 11834002, 12274070), Natural Science Foundation of Jiangsu Province (Grant No. BK20221451), Postgraduate Research \& Practice Innovation Program of Jiangsu Province (Grant Nos. KYCX24\_0362), and SEU Innovation Capability Enhancement Plan for Doctoral Students (Grant Nos. CXJH\_SEU 24035). Most calculations were done on the Big Data Computing Center of Southeast University.
\end{acknowledgments}

\bibliography{reftot}
\bibliographystyle{apsrev4-2}
\end{document}